\pdfoutput=1
\documentclass[aps,prl,twocolumn,longbibliography,superscriptaddress]{revtex4-2}
\usepackage{amsmath}

\usepackage{bm}
\usepackage{bbold}
\usepackage{amssymb}
\usepackage{graphicx}
\usepackage[ hidelinks = true, colorlinks = true, citecolor = blue, urlcolor = blue, linkcolor = blue]{hyperref}
\usepackage{verbatim}
\usepackage{color}
\usepackage[capitalise]{cleveref}
\usepackage[normalem]{ulem}
\usepackage{dsfont}
\usepackage{xcolor}
\usepackage{float}
\usepackage{placeins}

\allowdisplaybreaks[3]
\usepackage{dsfont}

\newcommand{\ats}[1]{\textcolor{red!80!black}}
\newcommand{\atstrike}[1]{%
  \ifmmode
    \textcolor{red!80!black}{\cancel{#1}}%
  \else
    \textcolor{red!80!black}{\sout{#1}}%
  \fi
}

\usepackage{accents}
\newlength{\dhatheight}

\begin{document}

\title{From coherent to fermionized microwave photons in a superconducting transmission line }

\author{Alberto Tabarelli de Fatis*}
\affiliation{Pitaevskii BEC Center, INO-CNR and Dipartimento di Fisica, Universit\`{a} di Trento, Via Sommarive 14, I-38123 Trento, Italy}
\email{a.tabarellidefatis-2@unitn.it}

\author{Stephanie Matern}
\affiliation{Pitaevskii BEC Center, INO-CNR and Dipartimento di Fisica, Universit\`{a} di Trento, Via Sommarive 14, I-38123 Trento, Italy}
\author{Gianluca Rastelli}
\affiliation{Pitaevskii BEC Center, INO-CNR and Dipartimento di Fisica, Universit\`{a} di Trento, Via Sommarive 14, I-38123 Trento, Italy}
\affiliation{INFN-TIFPA, Trento Institute for Fundamental Physics and Applications, Via Sommarive 14, I-38123 Trento, Italy}
\author{Iacopo Carusotto}
\affiliation{Pitaevskii BEC Center, INO-CNR and Dipartimento di Fisica, Universit\`{a} di Trento, Via Sommarive 14, I-38123 Trento, Italy}
\affiliation{INFN-TIFPA, Trento Institute for Fundamental Physics and Applications, Via Sommarive 14, I-38123 Trento, Italy}

\date{\today}

\begin{abstract}
We investigate superconducting transmission lines as a novel platform for realizing a quantum fluid of microwave photons in a propagating geometry. We anticipate that the strong photon-photon interactions provided by the intrinsic nonlinearity of Josephson junctions   are sufficient to enter a regime of strongly interacting photons for realistic parameters. To prove the potential of this platform, we show that a suitable tapering of the transmission line parameters allows for the adiabatic conversion of an incident coherent field into a cold Tonks-Girardeau gas of fermionized photons close to its ground state. Signatures of the strong correlations are visible in the correlation properties of the transmitted light.
\end{abstract}

\maketitle
Leveraging the effective photon-photon interactions stemming from the optical nonlinearity of the medium, quantum fluids of light have emerged as a promising new system to study many-body physics in novel regimes~\cite{Carusotto2013,Carusotto2020}. 
While most experiments so far are in a weak interaction regime, first steps towards strongly correlated regimes have been made with the observation of a Mott insulator state~\cite{ma2019dissipatively}, a spectroscopical study of a system of impenetrable photons~\cite{Fedorov:PRL2021}, and the realization of few-body fractional quantum Hall liquids~\cite{clark2020observation,Wang:Science2024}. These experiments have mostly been carried out in cavity configurations where losses are intrinsically needed to establish the steady state and to measure the quantum state of the photon gas.

On the other hand, systems with propagating geometries~\cite{GLORIEUX2025157} can be ideally operated in regimes with no losses where propagation is described by unitary evolution under the exchange of the role of spatial and temporal coordinates ~\cite{Lai:PRA1989,larre_propagation_2015}.
Most experiments so far have used visible light and focused on weakly interacting gases, with observations of superfluidity~\cite{michel2018superfluid}, thermalization and condensation phenomena~\cite{abuzarli2022nonequilibrium,fontaine2018observation}. At the same time, pioneering works have started exploring regimes with stronger nonlinearities, leading to the observation of marked antibunching and few-photon bound states~\cite{peyronel2012quantum,liang2018observation}. 
However, the observation of strongly quantum correlated many-body states has so far remained elusive.

Capitalizing on the development of circuit-QED technologies~\cite{Blais2021}, 
superconducting nonlinear transmission lines have experienced dramatic advances in the last years, e.g., in view of realizing quantum limited amplifiers -- the so called Traveling Wave Parametric Amplifiers (TWPA)~\cite{esposito2021perspective} -- and generating squeezed states of microwave fields~\cite{zhong2013squeezing}.
The design flexibility of state-of-the-art TWPAs and the strong nonlinearities provided by Josephson Junction (JJ) elements~\cite{Josephson1962} make these devices a natural platform to study and generate strongly correlated gases of microwave photons.

In this Letter we derive an effective quantum theory for the propagation of the microwave photon field in a superconducting transmission line in terms of a unitary Hamiltonian evolution, which provides a powerful tool for predicting quantum optical features.
As a proof of concept to demonstrate the potential of our approach, we show how a suitable tapering of the transmission line parameters allows to convert a monochromatic and fully coherent incident radiation into a one-dimensional Tonks-Girardeau (TG) gas of impenetrable photons~\cite{Tonks1936,Girardeau1960}. 

While the incident coherent field is straightforwardly obtained with commercially available microwave generators, the TG gas is a remarkable many-body state where strong optical nonlinearities make bosonic photons behave analogously to fermionic particles. In contrast to the strong size limitations of TG gases of ultracold atomic clouds~
\cite{Kinoshita2004,Paredes2004}, our scheme can generate a TG gas of an indefinitely large number of photons just by operating the device for a macroscopically long time, and at the same time offers direct access to the correlation functions of the TG gas.
Beyond antibunching, distinct signatures of the fermionized behavior~\cite{castin2004simple} of strongly interacting photons
appear as Friedel oscillations in the second-order correlation function. 
The correlated nature of the gas is also visible in the speed of the light-cone propagation of excitations after a sudden ramp~\cite{steinhauer2022analogue}, which follows the prediction of the Lieb-Liniger (LL) model~\cite{lieb1963exact1,lieb1963exact2}. 
Realization of the TG gas represents a first step towards the realization of many other strongly correlated states of light, such as a macroscopically-sized Mott insulator of photons. Beyond fundamental science, the resulting array of single-photon wavepackets opens remarkable opportunities for quantum technologies.
\begin{figure*}[t]
    \centering
    \includegraphics[width=\linewidth]{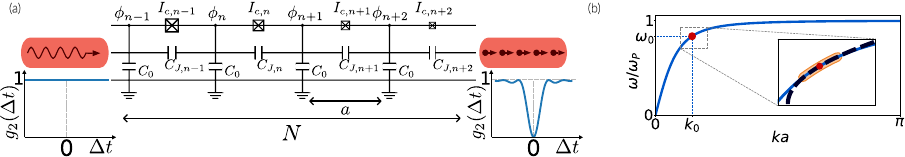}
    \caption{(a) Sketch of the transmission line model, composed of unit cells of length $a$ with a capacitor $C_0$ to ground and a series of $N$ JJs characterized by a critical current $I_{c,n}$ and a capacitance $C_{J,n}$, whose values are tapered along the propagation. The monochromatic coherent incident radiation, characterized by a flat second order photon correlation function $g_2(\Delta t)$, is converted into a gas of fermionized photons, with a $g_2(\Delta t)$ 
    showing antibunching and Friedel oscillations. (b) Sketch of the linear dispersion relation of the transmission line (solid blue line), together with the quadratic approximation around the carrier frequency $\omega_0$ (dashed black line). The incident monochromatic field is localized in frequency (red dot), whereas the fermionized photon gas after propagation has a finite spread in frequency (orange area).}
    \label{fig:pictorial}
\end{figure*}

We model a nonlinear superconducting transmission line based on JJs, as sketched in Fig.\,\hyperref[fig:pictorial]{1(a)}, with the Lagrangian~\cite{Rasmussen2021}
\begin{equation}
    \begin{split}
        L_{\text{circuit}} = & \frac{1}{2} \sum_nC_{0,n}\dot{
            \phi}_n^2 + \frac{1}{2}\sum_n C_{J,n}\left(\dot{\phi}_{n+1} - \dot{\phi}_n\right)^2 \\
            &+ \frac{\Phi_0}{2 \pi} \sum_{n} I_{c,n} \cos\left[\frac{2\pi}{\Phi_0}\left(\phi_{n+1} - \phi_n\right)\right].
            \label{eq:Lagrangian}
    \end{split}
\end{equation}
The flux variables $\phi_n$ are defined such that the superconducting phase difference across a JJ at position $n$ is $\varphi_n = {2\pi (\phi_{n+1}-\phi_n)}/{\Phi_0}$ with $\Phi_0 = {h}/{2e}$ the magnetic flux quantum. Furthermore, $I_{c,n}$ is the critical current of the JJ, $C_{J,n}$ its capacitance and $C_{0,n}$ the capacitance to ground.
Assuming a spatially homogeneous line, and restricting to quadratic terms in the fields in the Lagrangian, we obtain the linear dispersion relation of the transmission line, plotted in Fig.\,\hyperref[fig:pictorial]{1(b)}, 
\begin{equation}
    \omega^2 = \omega_P^2\frac{4c\sin^2\left(\frac{ka}{2}\right)}{1+4c\sin^2\left(\frac{ka}{2}\right)}\,.
\end{equation}
Here, we have assumed the unit cells to have constant length $a$, $c =C_J/C_0$, $\omega_P=1/ \sqrt{L_JC_J}$ is the plasma frequency and $L_J~=~ \Phi_0/(2\pi I_c)$ is the inductance of the JJ.

Nonlinear effects can be modeled assuming low field amplitude. This amounts to expanding the cosine in the Lagrangian of \cref{eq:Lagrangian} up to fourth order in the small parameter ${\phi_n}/{\Phi_0}$ and leads to the action $S=S_L+S_{NL}$, which consists of a linear and nonlinear part,  with
\begin{equation}
\begin{split}
    S_L =& \int dt \left[\frac{C_0}{2} \sum_n\dot{\phi}_n^2 + \frac{C_J}{2}\sum_n \left(\dot{\phi}_{n+1} - \dot{\phi}_n\right)^2-\right. \\ -&\left. \frac{1}{2L_J}\sum_n \left(\phi_{n+1} - \phi_n\right)^2 \right], \\
    S_{NL} =& \int dt  \left[\frac{1}{24L_J}\left(\frac{2\pi}{\Phi_0}\right)^2\sum_n \left(\phi_{n+1} - \phi_n\right)^4\right]. 
\end{split}
    \label{eq:nonlinear_action}
\end{equation}
Note that this form holds for generic transmission lines involving components whose current-phase relation is $I(\phi+\phi^*) = \phi/L_J- \gamma \phi^3$, such as Superconducting Quantum Interference Devices (SQUIDs)~\cite{kleiner2004superconducting}, Superconducting Nonlinear Asymmetric Inductive eLements (SNAILs)~\cite{frattini20173} operated in suitable regimes where three-wave mixing is negligible or other possible architectures \cite{ZhangW:2017}.

By introducing a continuous spatial coordinate $z=an$ along the propagation direction and expanding the field $\phi_n$ in Fourier components $\phi(z,t) = \int dk/(2\pi) \int d\omega/(2\pi) A(k,\omega)e^{i(kz-\omega t)}$, with $A(-k,-\omega) = A^*(k,\omega)$, we obtain for the linear action
\begin{equation}
\begin{split}
    S_L = & -\frac{C_0}{2a}\int \frac{dk}{2\pi} \int \frac{d\omega}{2\pi}|A(k,\omega)|^2 F(k,\omega).
\end{split}
\label{eq:linear_action}
\end{equation}
Here,  
\begin{equation}
    F(k,\omega) = -\omega^2\left[1+4c\sin^2\left(\frac{ka}{2}\right)\right]+4c\omega_P^2\sin^2\left(\frac{ka}{2}\right)
\end{equation}
describes the linear dispersion relation via $F(k,\omega)~=~0$. 
In the following, we restrict to circuit parameters inside of the superconducting phase of the Josephson transmission line, in order to avoid the insulating behaviour~\cite{PhysRevB.101.024518}. This roughly corresponds to working in regimes where the charging energy of the ground capacitance is comparable or smaller than the Josephson energy (see End Matter). 

Within the slowly-varying envelope approximation (SVEA), we can expand $F(k,\omega)$ around the carrier wavevector $k_0$ and frequency $\omega_0$ (see Fig.~\hyperref[fig:pictorial]{1(b)})
\begin{equation}
\begin{split}
    S_L \approx -\frac{C_0}{a}\left(\frac{\partial F}{\partial k}\right)_{(k_0,\omega_0)}
    \int \frac{d(\delta k)}{2\pi} \frac{d(\delta \omega)}{2\pi}|\tilde{A}(\delta k,\delta \omega)|^2 \times \\ \times \left(\delta k-\frac{1}{v_g}\delta \omega-\frac{D}{2}\delta \omega^2\right),
\end{split}
\end{equation}
with $\tilde{A}(\delta k= k-k_0, \delta \omega= \omega-\omega_0) = A(k,\omega)$.
The group velocity and curvature are evaluated at $(k_0,\omega_0)$ as $v_g~=~{\partial_k \omega}$ and $D =  {\partial_\omega^2 k}~=~-v_g^{-3}\,{\partial_k^2 \omega}$, and can be expressed in terms of derivatives of $F$ via explicit formulas (see Appendix).

Introducing the slowly varying field $\phi_0$ defined via $\phi(z,t) =  e^{i(k_0z-\omega_0 t)} \phi_0(z,t) + \text{c.c.}$, and computing the nonlinear term within the zeroth order SVEA, the total action reads
\begin{equation}
    \begin{split}
         S = \int d\tau \int d\zeta&\, \mathcal{C}\,\left[i\hbar \phi_0^*\partial_\tau \phi_0+i\hbar\frac{\bar{v}_g^2}{v_g}\phi_0^*\partial_\zeta\phi_0+\right.\\
         &\quad\left.+\frac{\hbar \bar{v}_g^3 D}{2}|\partial_\zeta\phi_0|^2-\frac{\tilde{g}}{2}|\phi_0|^4\right].
    \end{split}
    \label{eq:action}
\end{equation}
Here, we have exchanged the role of space and time via the change of variables $ \zeta = t \bar{v}_g$ and $\tau = {z}/{\bar{v}_g}$, with $\bar{v}_g$ an (arbitrarily chosen~\footnote{Introduction of $\bar{v}_g$ is convenient for mapping length into time, though the physical results are independent of it.}) reference velocity and $\mathcal{C}$ an overall normalization factor (see \cref{eq:norm_factor}), which plays a crucial role in the quantum theory since it modifies the commutation relations.

We note that \cref{eq:action} is the action of a nonlinear Schrödinger equation~\footnote{The dynamical variable $\phi_0$ is the only independent degree of freedom, as its conjugate momentum is proportional to $\phi_0^*$~\cite{cohen2024photons}. Therefore, in \cref{eq:action} only derivatives of $\phi_0$ appear.} with a drift velocity term, which can be removed by a change of variables $\zeta'~=~\zeta~-~\bar{v}_g^2\int_0^\tau v_g^{-1}(\tau')d\tau' $. 
In the following we consider that the circuit parameters slowly vary with the position $z$ along the transmission line, so that all parameters in our action slowly depend on the effective time $\tau$. 

From the action in \cref{eq:action} we derive an effective quantum theory  by imposing equal-$\tau$ commutation relations on the field $\hat{\phi}_0$ and its conjugate momentum~\cite{Lai:PRA1989,larre_propagation_2015}
\begin{equation}
    [\hat{\phi}_0(\zeta,\tau),\hat{\Pi}_{\phi_0}(\zeta',\tau)] = i\hbar \delta(\zeta-\zeta').
\end{equation}
Introducing the field operator
\begin{equation}
    \hat{\Psi}(\zeta,\tau) = \sqrt{\mathcal{C(\tau)}}\,\hat{\phi}_0(\zeta,\tau)
    \label{eq:psi}
\end{equation}
satisfying bosonic commutation relations $[\hat{\Psi}(\zeta,\tau),\hat{\Psi}^\dagger(\zeta',\tau)] = \delta(\zeta-\zeta')$, one obtains a Hamiltonian in the usual form of a one-dimensional Bose gas~\cite{castin2004simple} with contact interactions of strength $g(\tau)={\tilde{g}}/{\mathcal{C}(\tau)}$ and mass $m = -\hbar/(\bar{v}_g^3 D)$~\footnote{Higher order corrections to the linear dispersion relation beyond the SVEA would give a momentum dependent mass.}: 
\begin{equation}
    \hat{H} = \int_{-\infty}^{\infty} d\zeta\left[ -\frac{\hbar^2}{2m}\hat{\Psi}^\dagger (\zeta) \frac{\partial^2}{\partial \zeta^2} \hat{\Psi}(\zeta) + \frac{g}{2}\hat{\Psi}^{\dagger 2} (\zeta) \hat{\Psi}^2(\zeta) \right]\,.
    \label{eq:LLHamiltonian}
\end{equation}
For repulsive interactions the eigenstates of this LL Hamiltonian~\cite{lieb1963exact1,lieb1963exact2} are determined by the dimensionless interaction parameter $\gamma=m g/(\hbar^2\rho)$, where $\rho$ is proportional to the incident photon flux, as discussed later. As a peculiarity of our one-dimensional geometry, the impact of interactions increases for decreasing density $\rho$. In the strong interaction limit $\gamma \gg 1$~\footnote{Note that the expansion of the action \cref{eq:nonlinear_action} to the lowest nonlinear order is enough to describe the strongly interacting gas provided the field amplitude remains weak enough, i.e., $\phi/\Phi_0\ll1$. If needed, higher order terms would introduce multi-body interaction terms in the Bose gas Hamiltonian.}, the Bose gas approaches the $\gamma\to\infty$ TG regime, 
where it displays an effective fermionic behaviour~\cite{Girardeau1960}. 

Moving back to our specific superconducting transmission line model, we first note that according to the explicit formulas given in the End Matter, both the mass $m$ and the interaction strength $g$ are negative: modulo an overall sign in the Hamiltonian (which has no effect on the dynamics of an isolated system), the physics then recovers the one of a repulsively interacting gas. Secondly, the density $\rho$ of the Bose gas can be related to the incident photon flux by $\rho=\Phi_{ph}/\bar{v}_g$ (see End Matter): the $t$-translationally invariant stationary states under our monochromatic pump correspond to a spatially homogeneous gas in $\zeta$ with no limitation on its size. Furthermore, the two-time correlation functions $g_{1,2}(\Delta t = t_2-t_1, z=L)$ after propagation along a transmission line of length $L=Na$ are given by the corresponding two-position correlation functions $g_{1,2}(\Delta \zeta = \zeta_2-\zeta_1=\bar{v}_g(t_2-t_1), \tau = L/\bar{v}_g)$ of the Bose gas. Finally, while $g_{1,2}$ is ideally measured with single-photon detectors, schemes to also measure it for microwave photons using linear amplifiers and quadrature amplitude detectors have been developed~\cite{ PhysRevA.82.043804,PhysRevLett.106.243601,bozyigit2011antibunching}. 

According to the general theory of the TG gas~\cite{castin2004simple}, photon fermionization is apparent in the second-order photon correlations of the ground state
\begin{equation}\label{eq:g2TG}
    g_2^{TG}(\Delta t) =1-\left(\frac{\sin(\pi \Phi_{\text{ph}} \Delta t)}{\pi \Phi_{\text{ph}} \Delta t}\right)^2\,:
\end{equation}
a full antibunching is visible at short delays $\Delta t$, followed by Friedel oscillations at larger $\Delta t$. While the antibunching can be explained in terms of a two-photon blockade mechanism~\cite{peyronel2012quantum, PhysRevLett.106.243601} persisting at finite temperatures, the Friedel oscillations are a true many-body effect resulting from the quantum correlated nature of the ground state. Their many-body origin is apparent in the value of their frequency, which is set by the photon flux $\Phi_\text{ph}$ and corresponds to the Fermi wave vector. This is in contrast to resonance fluorescence \cite{CCT4}, where the oscillations in $g_2$ are set by the Rabi frequency proportional to the amplitude of the incident electric field.
On the other hand, the first order correlation function in the ground state monotonically decays at large time delays as a power law~\cite{castin2004simple}
\begin{equation}
    g_1^{TG}(\Delta t) \sim \frac{1}{\sqrt{\pi\Phi_{\text{ph}}|\Delta t|}}.
\end{equation}
\begin{figure*}[t]
    \centering
    \includegraphics[width=\textwidth]{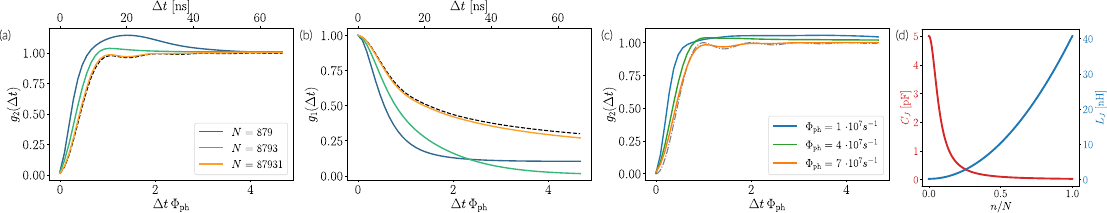}
    \caption{Second (a), (c) and first (b) order correlation function of the microwave field transmitted by the transmission line sketched in Fig. \hyperref[fig:pictorial]{1(a)}. The ramp of the circuit parameters is shown in (d), with a constant plasma frequency of $\omega_p=1/\sqrt{L_JC_J}=2\pi\cdot5$~GHz, a constant ground capacitance $C_0 = 30$~fF and an incident coherent field frequency $\omega_0/2\pi = 4.25$~GHz. The incident photon rate is  $\Phi_{\text{ph}}=7\cdot10^{7}~s^{-1}$ in (a) and (b) and varies according to the legend in (c). The number of cells $N$ of the transmission line varies according to the legend in (a) and (b), and is $N = 87931$ for (c). The dashed black line in (a) and (b) is the numerical ground state prediction computed numerically at the final value $\gamma\approx 33.2$, whereas the dashed-dotted line in (c) is the analytical TG result in the $\gamma=\infty$ limit.}
    \label{fig:slow_ramps}
\end{figure*}
In order to observe such quantum correlated many-body state, we propose to slowly vary the nonlinear properties of the transmission line so to adiabatically convert an incident monochromatic and fully coherent radiation of frequency $\omega_0$ into a TG gas of fermionized microwave photons close to the many-body ground state. In stark contrast to most experiments so far~\cite{GLORIEUX2025157,peyronel2012quantum, fontaine2018observation}, which involve fast variations of the optical nonlinearity upon entering the medium and thus lead to highly excited states, we propose to slowly vary the parameters of the transmission line in space so to adiabatically increase the LL parameter $\gamma$. 

As the LL model is gapless~\cite{lieb1963exact2} and the adiabatic theorem cannot be directly  applied~\cite{polkovnikov2008breakdown}, we employ $\tau$-dependent infinite tensor-network (TN) algorithms~\cite{PhysRevLett.98.070201, Orus2008} to numerically simulate the full many-body problem of the microwave propagation along the transmission line.
The specific adiabatic ramp in the temporal variable $\tau$ used in the numerics corresponds to the spatial variation of the circuit parameters along the transmission line sketched in Fig. \hyperref[fig:pictorial]{1(a)}: $C_0$ and $\omega_P=1/\sqrt{L_JC_J}$ are kept constant while $L_J$ and $C_J$ are changed as shown in Fig. \hyperref[fig:slow_ramps]{2(d)}. 

Note that our theory assumes from the outset that light is uni-directionally propagating along the transmission line and no reflections occur at its terminations. This can be ensured by impedance matching the transmission line with the input/output lines, e.g., with a suitable further tapering. Note that this latter tapering at the edges of the line can be done on a length scale of the order of the microwave wavelength, which imposes a much less severe constraint on the transmission line length than the many-body adiabaticity discussed above.

In \cref{fig:slow_ramps}, we show the correlation properties of the stationary microwave field that emerges after propagation along the tapered transmission line. In the figures, the propagation length is measured in terms of the number of cells $N$ in the transmission line, which does not depend on the physical size $a$ of the cell. The second order correlation function $g_2$ is shown in Fig.~\hyperref[fig:slow_ramps]{2(a)} for increasing values of the transmission line length but a fixed shape of the ramp profile.
The short distance antibunching due to the repulsive contact interactions is visible in all curves, while the Friedel oscillations typical of the TG ground state \cref{eq:g2TG} are well visible for the longest transmission line. This indicates that the gapless spectrum does not prevent the adiabatic evolution to approach the ground state provided the transmission line is sufficiently long. The good agreement with the exact ground state at the final $\gamma$ value is a quantitative verification of the efficiency of our adiabatic preparation scheme. Going deeper into the $\gamma\to\infty$ TG limit would only slightly increase the amplitude of Friedel oscillations (see End Matter).
The accurate preparation of the TG ground state is further illustrated in the plot of the first order correlation function shown in Fig.~\hyperref[fig:slow_ramps]{2(b)}: for the longest transmission line, $g_1(\Delta t)$ approaches the power-law scaling of the TG ground state, again in good agreement with the exact ground state. 
Our specific simulations show that a number of cells around $N\approx8.8\cdot10^4$ is sufficient to generate a TG gas close to the ground state. While this value is somehow larger than existing waveguide devices~\cite{ergul2017spatial}, much longer devices in the range of interest are routinely used for metrological applications~\cite{Behr_2012}.

In addition to fabricating longer transmission lines, the efficiency of the adiabatic ramp can also be improved by increasing the incident photon flux, as shown in Fig.~\hyperref[fig:slow_ramps]{2(c)}, which softens the adiabaticity constraint on the length of the transmission line.
This trick however faces limitations: the LL parameter $\gamma$ decreases with $\Phi_{\text{ph}}$ but has to remain in the $\gamma\gg 1$ strong interaction regime. On top of this, increasing $\Phi_{\text{ph}}$ also increases the spread in the frequency distribution of the TG ground state, so the SVEA might eventually break down. A detailed discussion of the validity of our approximations for realistic parameters is given in the End Matter.

A further way to reduce the required length of the transmission line utilizes an overall scaling invariance of the Hamiltonian. The temporal variable $\tau$ can be reduced by a factor $\alpha$ by realizing a rescaled Hamiltonian $\alpha H$, which can be achieved by uniformly increasing $L_J(\tau)$ by $\alpha$ while keeping $C_0$, $C_j(\tau)$ and $r$ constant. 
Further improvements may be obtained by designing  more complex transmission lines based on SQUIDs or SNAILs or opening a gap in the linear dispersion via  a periodic modulation of the transmission line parameters.  

\begin{figure}[h]
    \centering
    \includegraphics[width=\linewidth]{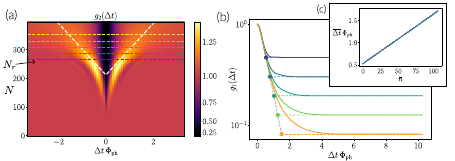}
    \caption{Propagation of correlations after a moderately fast ramp during $N_r=263$ cells (same ramp parameters as in Fig. \hyperref[fig:slow_ramps]{2(a)}) followed by a uniform region with constant transmission line parameters. (a) 
    Color plot of the second order photon correlation function $g_2(\Delta t)$ as a function of the propagation length in units of unit cells $N$. 
    The color of the dashed lines correspond to 
    the different curves in (b), the arrow indicates the end of the ramp. The dashed white line indicates motion at the expected LL speed $2c_{LL}$. (b) Plots of the first order correlation function $g_1(\Delta t)$ at different propagation lengths $n$. 
    The gray dashed line is an exponential fit of the short time decay of the longest-$\bar{z}$ curve; the horizontal dashed lines are the large $\Delta t$ asymptotes. 
    For each propagation length $\bar{n}$, 
    the intersection between these two lines (colored points) defines $\overline{\Delta t}(\bar{n})$. 
    The excellent agreement of the numerically extracted $\overline{\Delta t}(\bar{n})$ 
    (dots) with the theoretically expected light-cone spreading at the LL speed $c_{LL}$ (solid line) is shown in  (c).
    }
    \label{fig:fast_ramps}
\end{figure}

As an alternative strategy, it is interesting to note that signatures of the quantum correlated nature of the strongly interacting photon gas are already visible for moderately short transmission lines. In \cref{fig:fast_ramps} we show the correlation properties of the transmitted microwaves for a transmission line configuration displaying a relatively fast spatial ramp over $N_r$ cells followed by a uniform region of varying length $\bar{n}=N-N_r$ 
(parameters in the caption). On top of the always visible antibunching at small $\Delta t$, an additional positive bump develops in $g_2(\Delta t)$ after the ramp, whose position shifts to larger $\Delta t$ for growing $\bar{n}$, as shown in Fig. \hyperref[fig:fast_ramps]{3(a)}. 

A related interesting dynamics is visible in the dynamics of the first order correlation function $g_1(\Delta t)$ during propagation  shown in Fig.~\hyperref[fig:fast_ramps]{3(b)}. 
Analogously to the recent atomic experiment of~\cite{langen2013local}, immediately after the fast ramp the $g_1(\Delta t)$ correlation function retains memory of the initial coherent state 
, but its late-time value at large $\Delta t$ progressively drops during propagation along the uniform part of the transmission line. As done in~\cite{langen2013local}, we model $g_1(\Delta t)$ at small $\Delta t$  with an exponential decay $g_1(\Delta t)\sim e^{-\Gamma \Delta t}$ and a $\Delta t$-independent (but $n$-
dependent) value at large $\Delta t$. 
For each value of $n$,  
the transition point between the exponential and constant region is computed by locating in Fig.~\hyperref[fig:fast_ramps]{3(b)} the intersection point between the horizontal lines at the asymptotic constant values 
and the short-distance exponential decay. 

As $\bar{n}$ increases, this transition point $\overline{\Delta t}$ between the two regions is expected to move outwards according to the typical light cone spreading~\cite{calabrese2006time} $\bar{v}_g^2\partial_{\bar{z}} \overline{\Delta t}=2 c_{LL}$ with $\bar{z}=a\bar{n}$. In contrast to the weakly interacting fluids of recent atomic~\cite{langen2013local} or photonic~\cite{steinhauer2022analogue} experiments, however, the light-cone expansion speed $c_{LL}$ 
is here determined the many-body LL model~\cite{lieb1963exact1,lieb1963exact2}.
As it is shown in the inset of Fig.~\hyperref[fig:fast_ramps]{3(c)}, the value of the speed extracted from the numerical data differs from the expected theoretical LL value by only 0.5\%.
`To summarize, we have developed a theory for the propagation of microwave photons in a strongly nonlinear superconducting transmission line. Exchanging the role of space and time, propagation along the line can be described in terms of the Hamiltonian evolution in the form of an interacting one-dimensional Bose gas. Using this theory we propose a scheme to generate a Tonks-Girardeau gas of impenetrable photons with effective fermionic properties. Using a suitable tapering of the transmission line, an adiabatic preparation of the ground state is anticipated; for faster ramps, signatures of strong quantum correlations are visible in the light-cone propagation of excitations at a speed determined by the Lieb-Liniger model of one-dimensional interacting bosons. As a key advantage with respect to cavity configurations, the continuous-wave operation of our propagating scheme allows to overcome the limitations in the spatial size of the system and, thus, opens the way to realizing gases of a macroscopic number of photons.

As natural next steps, we plan to address more complex transmission line geometries with an optimized dispersion relation and more sophisticated SQUID or SNAIL elements. This will allow to reinforce the adiabaticity condition and, in this way, reduce the requirement on the transmission line length. Even though our discussion was focused on a specific circuit-QED setup, our predictions are straightforwardly extended to generic strongly nonlinear waveguide systems, both in the microwave and optical regimes, including kinetic inductance devices~\cite{PhysRevApplied.21.024011}, polaritons in solid-state systems~\cite{Walker:APL2013,suarez2021enhancement,rosenberg2018strongly} and Rydberg-EIT configurations in atomic gases~\cite{peyronel2012quantum,liang2018observation}. As future perspectives, our approach can be used to explore degenerate gases of multiphoton bound states under attractive interactions~\cite{horvath2025observing} and strongly interacting Bose gases in multi-waveguide geometries in the presence of synthetic magnetic fields~\cite{oliver2023artificial},\nocite{cohen2024photons} so to realize, for instance, two-dimensional fractional quantum Hall liquids of light.

\paragraph{Acknowledgements}
We acknowledge useful discussions with Felix Ahrens, Giulio Cappelli, Nicol\`o Crescini, Quentin Glorieux, Christian Johansen, Pierre-\'Elie Larr\'e, Federica Mantegazzini, Pasquale Scarlino.
This work was supported by the Provincia Autonoma di Trento; by the Q@TN Initiative; by the National Quantum Science and Technology Institute through the PNRR MUR Project under Grant PE0000023-NQSTI, co-funded by the European Union -- NextGeneration EU; by the Deutsche Forschungsgemeinschaft (DFG, German Research Foundation) via the Research Unit FOR 5688 (Project No. 521530974).


\bibliography{references}

\clearpage

\appendix

\onecolumngrid
\section{Appendix}
\subsection{Dependence of Bose gas properties on circuit parameters}
The explicit form of the physical quantities in \cref{eq:action} and \cref{eq:LLHamiltonian} are the following.
The overall normalization of the action is
\begin{equation}
    \mathcal{C} = \frac{r\sqrt{-(1+4c)r^2+4c}}{\hbar c L_J \bar{v}_g}\,.
    \label{eq:norm_factor}
\end{equation}
The group velocity is
\begin{equation}
    v_g = \frac{\omega_P}{2}a\sqrt{-(1+4c)r^2+4c}(1-r^2)\,.
    \label{eq:vg_circuit}
\end{equation}
The mass and interaction constant of the one-dimensional Bose gas are
\begin{equation}
\begin{aligned}
    m &=-\frac{\omega_P^2}{2}\frac{a\hbar}{\bar{v}_g^3r}\frac{\left[-(1+4c)r^2+4c\right]^{3/2}(1-r^2)}{1+12c-\frac{2r^2}{1-r^2}}
    \label{eq:m_circuit} \\
    g &= -\left(\frac{2\pi}{\Phi_0}\right)^2\frac{\hbar^2\bar{v}_g^2 L_J}{2a}\frac{r^2}{\left[-(1+4c)r^2+4c\right](1-r^2)^2}\,.
\end{aligned}
\end{equation}
Here, we have used the short-hands $r = \omega_0/\omega_P$, $\omega_P=1/\sqrt{L_JC_J}$, $c=C_J/C_0$.

The density of the Bose gas can be derived from the incoming photon flux $\Phi_{\text{ph}}$ by noting that the energy per unit length transported by incoming photons traveling at $v_g$ is $\mathcal{E}_1=\hbar\omega_0\Phi_{\text{ph}}/v_g$. This must equal the energy density computed from \cref{eq:Lagrangian} for a plane wave in the linear regime. Imposing a plane-wave form of $\phi_n$, inserting it into the Hamiltonian corresponding to the Lagrangian of \cref{eq:Lagrangian} and expressing the resulting energy in terms of the expectation value of the Bose field $\hat{\Psi}$ of \cref{eq:psi}, we get $\mathcal{E}_2=\hbar \omega_0 \bar{v}_g|\Psi|^2/v_g $, which immediately implies that $\rho = |\Psi|^2 = \Phi_\text{ph}/\bar{v}_g$.

\subsection{Validity of approximations}
\begin{figure}[t]
    \centering
    \includegraphics[width=0.6\linewidth]{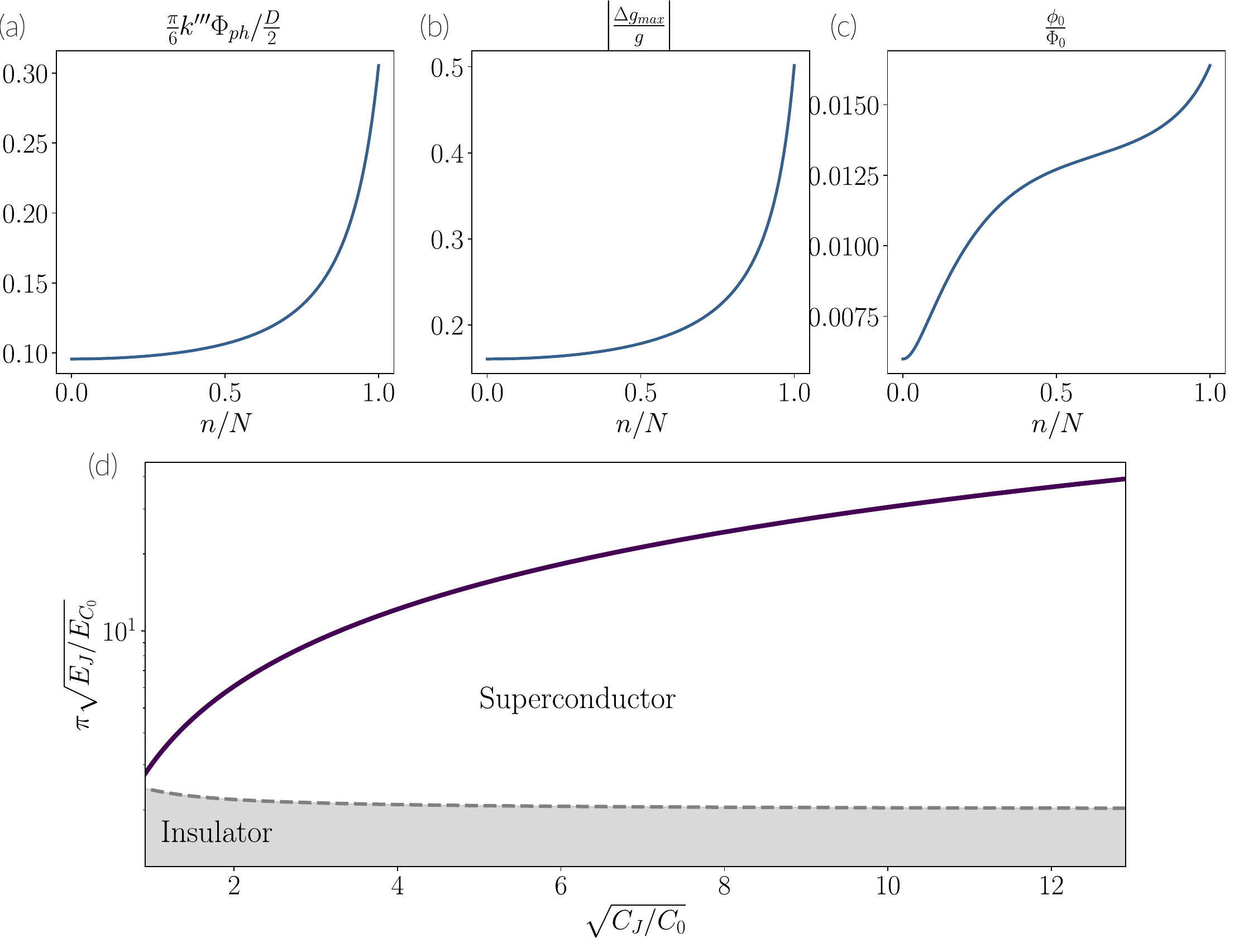}
    \caption{Validity of the approximations. First correction to parabolic dispersion (a), variation of the contact interaction parameter (b), and field amplitude (c). Panel (d) shows the phase diagram of the transmission line as in \cite{PhysRevB.101.024518}, the solid line represents the chosen parameters for the tapered transmission line. Parameters as in Fig. \ref{fig:slow_ramps}(a,b) in the main text. 
    }
    \label{fig:ramp_profile_approx}
\end{figure}

As explained above, our approach is based on expanding the linear dispersion relation up to second order around the carrier at $(k_0,\omega_0)$. This SVEA approximation must hold throughout the full propagation, and in particular in the TG regime. As the spread in frequency of the TG ground state is given by $\Delta \omega= \pi \Phi_{\text{ph}}$, the condition for the third order dispersion term to be negligible reads 
\begin{equation}
    \frac{1}{2}\left|\frac{\partial^2 k}{\partial \omega^2}\right|_{k_0} \gg \frac{\pi}{6}\left|\frac{\partial^3 k}{\partial \omega^3}\right|_{k_0} \Phi_{\text{ph}}\,.
\end{equation}
The ratio of these two quantities is plotted in Fig. \hyperref[fig:ramp_profile_approx]{4(a)}: the fact that it is indeed much smaller than one for the chosen parameters validates our approximation.

In Fig. \hyperref[fig:ramp_profile_approx]{4(b)} we plot the maximum relative variation of the interaction constant $g$ in the spectral region $\Delta \omega$ of interest: while the variation is moderate but significant, it is not expected to impact the validity of our findings, since a slight variation of the interaction strength does not affect the overall many-body properties of the TG gas.

Moreover, in Fig. \hyperref[fig:ramp_profile_approx]{4(c)} we show the field amplitude $\phi_0/\Phi_0$: its small value supports our expansion of the cosine in \cref{eq:Lagrangian} up to the quartic term only.

Finally, in Fig. \hyperref[fig:ramp_profile_approx]{4(d)} we show the phase diagram of the transmission line as a function of the ratio of the two capacitances, and of the Josephson $E_J = \Phi_0^2/(4\pi^2L_J)$ and charging $E_{C_0}=(2e)^2/C_0$ energies, as computed in \cite{PhysRevB.101.024518}, and show we always lie on the superconducting side along the whole transmission line.

\subsection{Effect of a finite interaction constant $\gamma$}

\begin{figure}[h]
    \centering
    \includegraphics[width=0.6\linewidth]{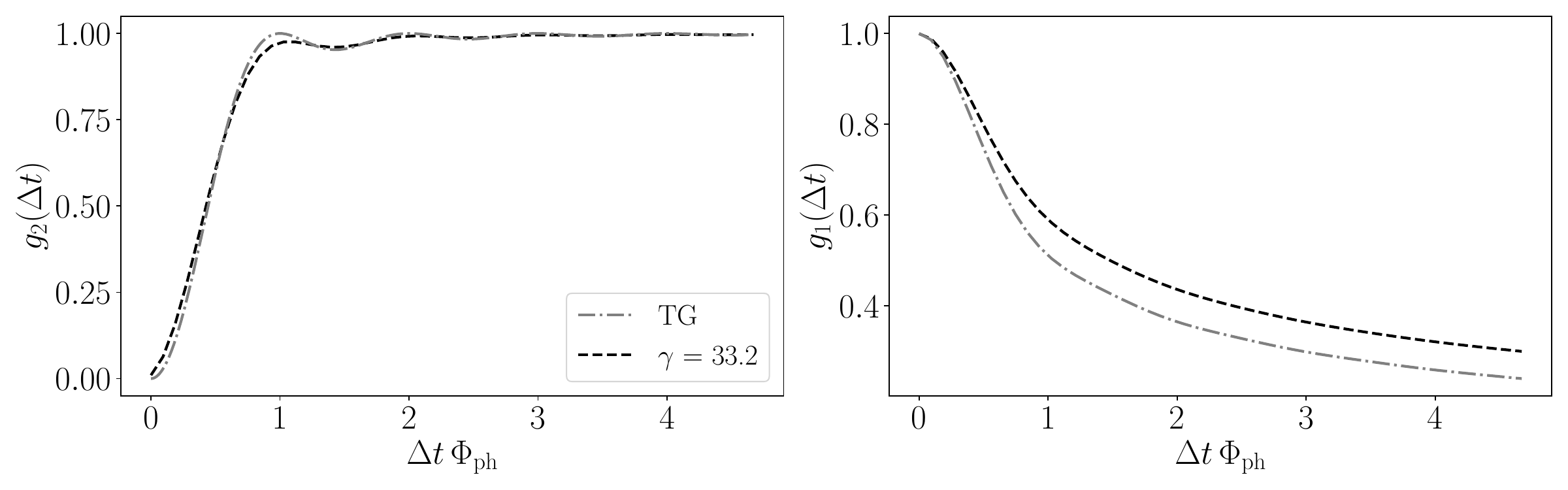}
    \caption{First (left) and second (right) order correlation functions in the TG limit (gray dashed-dotted line) and for the finite $\gamma$$\approx 33.2$ (black dashed line) used in the text.}
    \label{fig:finite_gamma}
\end{figure}

In Fig.~\hyperref[fig:slow_ramps]{2} of the main text, we have shown the numerically calculated correlation functions of the field after the ramp. The ground state correlation functions at finite $\gamma$ was computed by means of an imaginary time evolution on the infinite-MPS ansatz~\cite{Orus2008}. In \cref{fig:finite_gamma}, the result for a value $\gamma \approx 33.2$ corresponding to the end of the ramp is compared to the predictions for the TG gas at $\gamma\to\infty$. The good agreement of the curves show that for this value of $\gamma$ we are already well into the TG regime. The only qualitative effect of the finite $\gamma$ is a slight reduction of the amplitude of the Friedel oscillations.

\subsection{Comment on experimental feasibility}
In this section, we comment on the experimental feasibility of the proposed experiment.\\

{\sl \bf Sampling rate for $g_{1,2}(\Delta t)$.} 
Firstly, fast resolution around $2$-$3$~ns is required to resolve features in the pair correlation functions shown in the main text. While experimental observations so far~\cite{PhysRevA.82.043804, PhysRevLett.106.243601} report a resolution of around $10$~ns, the slight increase of the sampling rate that is needed to resolve these features is well within the possibility of state-of-the-art electronics. This comes at the price of increasing the noise bandwidth. The ensuing reduction of the signal to noise ratio can be compensated with a longer sampling time, allowing to detect small features like the Friedel oscillations.\\
Alternatively, one may also consider increasing the characteristic time-scale of the correlation features, which is inversely proportional to the incident photon flux. As this comes with the price of a smaller characteristic energy, a compromise would need to be taken with the device length so to preserve adiabaticity.\\

{\sl \bf Range for the plasma frequency.} 
While the proposed value of $5\,\text{GHz}$ for the plasma frequency is lower than the typical parameters for Josephson junctions (JJs), high-$L_J$ junctions with low plasma frequencies in this range have been fabricated in state-of-the-art devices~\cite{monroe2021optical,PhysRevApplied.14.064006,manset2025hyperinductancebasedstackedjosephson}. Moreover, replacing simple JJs with more sophisticated geometries such as SQUIDs or SNAILs~\cite{kleiner2004superconducting,frattini20173} enables a further reduction of the effective plasma frequency by exploiting their flux tunability.\\

{\sl \bf Number of cells.}
We note that for the specific ramp we considered, a transmission line with a number $N\approx8.8\cdot10^4$ of cells is largely sufficient to adiabatically generate a Tonks-Girardeau gas close to its ground state. While this value is around an order of magnitude larger than the currently existing waveguide devices \cite{ergul2017spatial}, it is well comparable to the one of Josephson arrays for metrological applications~\cite{Behr_2012}. Furthermore, the constraint on the device length can be loosened by further reducing the effective plasma frequency, as explained in the main text, or by engineering the linear dispersion relation in order to reduce the effective mass $m$ of the Bose gas. This can be done, for example, by considering a two-site unit cell and periodically modulating the junction parameters~\cite{PhysRevX.10.021021}. \\

{\sl \bf Effect of the photon losses and disorder.}
In an actual experimental setup, losses, though small, are generally unavoidable. Their main effect in our fermionized system is to heat up the gas, due to the fact that losing a fermion creates a hole excitation in the Fermi sphere. An upper bound on the excitation energy resulting from a fractional intensity drop of $\beta = 1- I_f/I_i$, can be estimated assuming that all particles are lost at the center of the Fermi sphere. This gives $\Delta E/E_0 =3\beta/(1-\beta)^2$. From this we expect that a typical temperature of $T\sim0.2\,T_F$, for which Friedel oscillations are damped by a factor of 2, results from a loss of $\beta \approx 6\%$. We note that this is a worst-case estimate, as losses occurring during the initial part of the propagation do not actually result in gas heating, due to the coherent nature of the photon gas in the early stages of propagation.\\ 
Finally, we expect that a small disorder along the transmission line does not result in a sizeable heating of the gas, as disorder corresponds to a small shaking of the gas parameters on extremely fast timescales, which is therefore averaged out.

\subsection{Derivation of the Schrödinger action}
The linear part of the action presented in the paper
\begin{equation}
    S_L = \int dt \left[\frac{C_0}{2} \sum_n\dot{\phi}_n^2 + \frac{C_J}{2}\sum_n \left(\dot{\phi}_{n+1} - \dot{\phi}_n\right)^2- \frac{1}{2L_J}\sum_n \left(\phi_{n+1} - \phi_n\right)^2 \right],
    \label{eq:nonlinear_action}
\end{equation}
is approximated within the Slowly-Varying Envelope Approximation (SVEA) by Taylor expanding $F(k,\omega)$ around a point $(k_0,\omega_0)$ on the dispersion relation, for which $F(k_0,\omega_0)$ = 0. We expand $F(k,\omega)$ up to second order
and we exploit the symmetry of $F(k,\omega)$ under the exchange $(k,\omega)\to (-k,-\omega)$ to obtain
\begin{equation}
    \int \! dk\, d\omega \,|A(k,\omega)|^2 F(k,\omega) \approx  
     2 \int d(\delta k) d(\delta \omega) |\tilde{A}(\delta k,\delta\omega)|^2\tilde{F}(\delta k,\delta\omega)
\end{equation}
with $\tilde{A}(\delta k= k-k_0, \delta \omega= \omega-\omega_0) = A(k,\omega)$ and
\begin{equation}
    \tilde{F}(\delta k, \delta\omega) = \frac{\partial F}{\partial k}\delta k+\frac{\partial F}{\partial \omega}\delta \omega+\frac{1}{2}\frac{\partial^2 F}{\partial k^2}\delta k^2+ \frac{1}{2}\frac{\partial^2 F}{\partial \omega^2}\delta \omega^2 + \frac{\partial^2 F}{\partial k \partial \omega}\delta\omega \delta k.
\end{equation}
We can now replace in all but the first term the dispersion relation $\delta k = \delta \omega/v_g+\delta \omega^2D/2$ obtained from the expansion around $(k_0,\omega_0)$, getting
    \begin{equation}
        \int \!dk\, d\omega \,|A(k,\omega)|^2 F(k,\omega) \approx 2\left(\frac{\partial F}{\partial k}\right) \int \! d(\delta k)\, d(\delta \omega)\,|\tilde{A}(\delta k,\delta \omega)|^2\left(\delta k-\frac{1}{v_g}\delta \omega-\frac{D}{2}\delta \omega^2\right).
    \end{equation}
Interestingly, both the group velocity $v_g$ and the dispersion coefficient $D$ can be computed from $F$ even if the explicit form of $\omega(k)$ is not known, as
\begin{equation}
    v_g = \frac{\partial \omega}{\partial k}\Big\lvert_{(k_0,\omega_0)}= -\frac{\frac{\partial F}{\partial k}\Big\lvert_{(k_0,\omega_0)}}{\frac{\partial F}{\partial \omega}\Big\lvert_{(k_0,\omega_0)}}  
    \label{eq:vg_F}
\end{equation}
and
\begin{equation}
        D =\frac{\partial^2 k}{\partial \omega^2} = -\frac{1}{v_g^3}\frac{\partial^2 \omega}{\partial k^2}
        =-\left(\frac{\partial F}{\partial k}\right)^{-3}\left[\frac{\partial^2 F}{\partial \omega^2}\left(\frac{\partial F}{\partial k}\right)^2\right.
        \left.+\frac{\partial^2 F}{\partial k^2}\left(\frac{\partial F}{\partial \omega}\right)^2-2\frac{\partial F}{\partial \omega}\frac{\partial F}{\partial k}\frac{\partial^2 F}{\partial \omega \partial k}\right]
        ,
    \label{eq:D_F}
\end{equation}
where all derivatives are taken at $(k_0,\omega_0)$.

Defining the envelope field $\phi_0$ as
\begin{equation}
\begin{split}
    \phi(z,t) & = e^{i(k_0z-\omega_0 t)}\iint
    \frac{d\delta k}{2\pi} \frac{d\delta\omega}{2\pi} \tilde{A}(\delta k,\delta\omega)e^{i(\delta k z-\delta\omega t)} + c.c.\\
    & = e^{i(k_0 z-\omega_0 t)} \phi_0 (z,t) + e^{-i(k_0z-\omega_0 t)} \phi_0^*(z,t)\,,
\end{split}
\label{eq:phi_0}
\end{equation}
we are led to the Schrödinger action
\begin{equation}
         S = \int d\tau \int d\zeta\, \mathcal{C}\,\left[i\hbar \phi_0^*\partial_\tau \phi_0+i\hbar\frac{\bar{v}_g^2}{v_g}\phi_0^*\partial_\zeta\phi_0+\frac{\hbar \bar{v}_g^3 D}{2}|\partial_\zeta\phi_0|^2-\frac{\tilde{g}}{2}|\phi_0|^4\right].
    \label{eq:action}
\end{equation}

\subsection{Numerical Methods}
For the numerical calculations presented in the main text, we map the problem onto a lattice model of spacing $l$ by discretize the variable $\zeta$ as $\zeta_j = l\, j$.
For small enough $l$, we can approximate the Laplacian with finite differences and reformulate the system in terms of the Bose-Hubbard Hamiltonian
\begin{equation}\label{eq:LatticeHam}
    \hat{H}(t)  = -J \sum_j\left(\hat{a}_j^\dagger \hat{a}_{j+1}+
    \hat{a}_{j+1}^\dagger\hat{a}_j\right) +2J\sum_j\hat{n}_j+\frac{g(t)}{2}\sum_j\hat{n}_j(\hat{n}_j-1)
\end{equation}
where the on-site operators $\hat{a}_j = \sqrt{l}\,\hat{\Psi}(\zeta_j)$ satisfy Bose commutation rules $[\hat{a}_j,\hat{a}_{j'}^\dagger] = \delta_{j,j'}$. In order to best approximate the continuous nature of our physical system, we choose $J = \hbar^2/(2ml^2)$ and $g = g_0/l$.
The main limitation of the lattice model is that the kinetic energy dispersion is no longer quadratic, but has the typical cosine shape of lattice models 
\begin{equation}
    \epsilon(k) = 2J (1-\cos kl)
\end{equation}

While this approximation is able to accurately capture the low energy, long-wavelength physics at $k l\ll 1$, the lattice model fails to describe the continuum physics on shorter length scales for which $k l\gtrsim 1$.

In order to simulate this strongly correlated one-dimensional system an infinite-MPS ansatz~\cite{PhysRevLett.98.070201,Orus2008} was employed to describe an infinite and translationally invariant lattice in the $\zeta$ variable, corresponding to a stationary, time independent solution in the physical time $t$. The dynamics was simulated with the time evolution by block-decimation (TEBD) algorithm, which was implemented with the ITensor library~\cite{10.21468/SciPostPhysCodeb.4}. The system is initialized as a coherent state with a given density $\rho$, and the cutoff of the local bosonic Hilbert space on each site is chosen such that the occupation of the neglected states at large photon number is much smaller than one. This automatically ensures that the lattice spacing is much smaller than the characteristic length scales of the photon gas, both in the initial weakly-interacting and in the final TG states.

\subsection{Other Correlation Functions}
Beyond the first and second order coherences presented in the main text, other correlation functions of the transmitted microwave field can be compared to the Tonks-Girardeau ground state predictions. In particular, we focus on the third order coherence function:
\begin{equation}
    g_3(\Delta t_{1}, \Delta t_{2}) = \frac{\langle \hat{\Psi}^\dagger(\Delta t_{1}) \hat{\Psi}^\dagger(\Delta t_{2}) \hat{\Psi}^\dagger(0)  \hat{\Psi}(0) \hat{\Psi}(\Delta t_{2}) \hat{\Psi}(\Delta t_{1})\rangle}{\rho^3}
\end{equation}
as well as a mixed correlator:
\begin{equation}
    g_m(\Delta t_{1}, \Delta t_{2}) = \frac{\text{Re}\left[\langle \hat{\Psi}^\dagger(0) \hat{\Psi}^\dagger(\Delta t_{1})  \hat{\Psi}(\Delta t_{2}) \hat{\Psi}(0)\rangle\right]}{\rho^2}
\end{equation}
where $\rho = \langle \hat{\Psi}^\dagger(0)  \hat{\Psi}(0) \rangle$. We assume that all the fields are evaluated after the propagation, so that the propagation length $z=L=Na$ is fixed to its value at the end of the ramp.

\begin{figure}[t]
    \centering
    \includegraphics[width=0.75\textwidth]{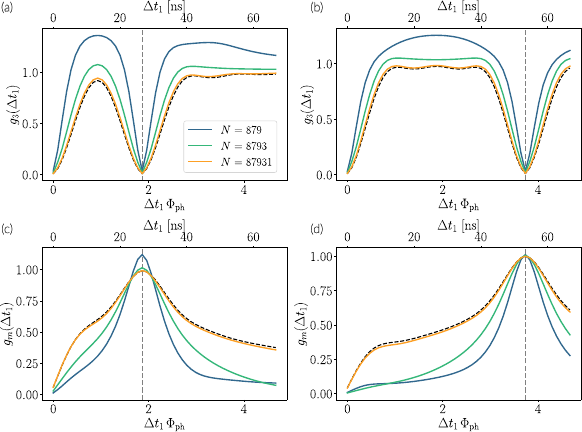}
    \caption{Third order coherence $g_3(\Delta t_1,\Delta t_2)$ (a), (b) and mixed correlator $g_m(\Delta t_1,\Delta t_2)$ (c), (d), of the microwave field transmitted by the transmission line as a function of the first time-delay $\Delta t_1$. The value of the second time delay is instead kept fixed and corresponds to the gray dashed line, with $\Delta t_2 \approx 1.88~$ns (a), (c) and $\Delta t_2 \approx 3.74~$ns (b), (d). The parameters are the same used in Fig.2 of the main text, i.e. with a constant plasma frequency of $\omega_p=1/\sqrt{L_JC_J}=2\pi\cdot5$~GHz, a constant ground capacitance $C_0 = 30$~fF, and an incident coherent field of frequency $\omega_0/2\pi = 4.25$~GHz and photon rate $\Phi_{\text{ph}}=7\cdot10^{7}~s^{-1}$. The different colors correspond to different lengths of the transmission line, as indicated in the legend. The dashed black line is the numerical ground state prediction computed numerically at the final value $\gamma\approx 33.2$.}
    \label{fig:correlators}
\end{figure}

\cref{fig:correlators} shows the expected value of the two correlators after the propagation in the transmission line, for the same parameters used in the main text and different transmission line lengths. Remarkably, the antibunching is also observable in the third order coherence function $g_3$, when $\Delta t_1 = 0, \Delta t_2$. As it happened for the first and second coherence functions shown in the main text, some deviations are visible for short ramps (blue and green), but for a very long ramp (yellow) the correlation functions of the transmitted field perfectly recover the very non-trivial features predicted for the ground state (black dashed line). This further validates our claim that the output field is indeed a Tonks-Girardeau gas close to its ground state.

\end{document}